\documentclass[twocolumn,showpacs,preprintnumbers,amsmath,amssymb]{revtex4}

\usepackage{graphicx}
\usepackage{dcolumn}
\usepackage{bm}

\begin{document}

\title{Continuously decoupling a Hadamard quantum gate from independent classes of errors}

\author{F. F. Fanchini}
 \email{felipe@ifsc.usp.br}
 
\author{J. E. M. Hornos}
 
\author{R. d. J. Napolitano}
\affiliation{Instituto de F\'{\i}sica de S\~{a}o Carlos, Universidade de S\~{a}o Paulo, Caixa Postal 369, 13560-970, S\~{a}o Carlos, SP, Brazil}

\date{\today}

\begin{abstract}
We consider protecting a Hadamard operation from independent dephasing, bit flipping, and dissipation. These environment-induced errors are represented by three uncorrelated reservoirs of thermalized bosons and we show that the protection is achievable through continuous dynamical decoupling. We find that, to decouple the Hadamard evolution from the environmental influence, we need a control field of higher frequency if the boson spectral density is super-ohmic than if it is ohmic. We also study the relevance of bit flipping and dissipation to the gate fidelity when it is protected from dephasing, showing how robust this partial protection is against these other perturbations. Finally, we calculate an efficient field arrangement capable of protecting simultaneously the gate operation from these three error classes.

\end{abstract}


\maketitle
\section{INTRODUCTION}
A quantum computer, when finally built, will be more efficient
than current classical computers to solve certain kinds of problems
\cite{deutsch92,shor94}. For instance, to find the prime factors of a large integer $N$ of $L$ bits, a quantum algorithm will take a time that is a polynomial function of $L$, while any classical procedure known at present takes much longer \cite{shor97}. Quantum information processing generally takes advantage of the inherent parallelism exhibited by unitary operations on quantum-state superpositions. The terms of these linear combinations are tensor products of quantum bits, or ``qubits'' \cite{schumacher95}. A qubit is, in analogy with the ``bit'' of classical computing, the elementary unit of quantum information, represented
by any state of a two-state system. Thus, if the two-dimensional Hilbert
space for this system has an orthonormal basis given by the kets $\left|0\right\rangle$ and $\left|1\right\rangle$, then a qubit is any linear combination of these basis states: $\alpha\left|0\right\rangle + \beta\left|1\right\rangle$. There is, therefore, a continuum of different qubits, while there are only two distinct classical bits: $0$ and $1$.

The theory of quantum information processing is based on the
preparation, unitary transformations, and measurements of superposed tensor products of qubits \cite{nielsen00}. The initial stage, characterized by the preparation of the computation input, and the last stage, corresponding to reading the output, consist of
measuring certain observables. Generating the initial superposition of states
and processing quantum information are, therefore, to be achieved
by performing unitary operations that cause the required interference and entanglement of
the input qubits. However, the actual quantum evolution of any system is not strictly
unitary, because the system cannot be completely isolated from the universe.
The consequent decay of the quantum-state purity is a manifestation of the celebrated
phenomenon of decoherence \cite{zurek91}.

There are three classes of strategic devices proposed to counteract the deleterious and unavoidable effects of decoherence: quantum error correcting codes (QECC) \cite{shor95, steane96, calderbank96, amsteane96, msbyrd04}, decoherence-free subspaces (DFS) \cite{zanardi97, lidar98, eknill00, pemfm05}, and dynamical decoupling (DD) \cite{viola98, viola99, viola99b, viola04}. Each one of these strategies must satisfy proper conditions for succeeding. For example, for a QECC, it is necessary to use auxiliary qubits and, generally, a low error rate is assumed so that just one qubit is perturbed during the protocol execution. For a DFS strategy, besides auxiliary qubits, a special environment symmetry that gives the direction for a suitable decoherence-free qubit encoding is also necessary. Finally, the DD requires, in principle, only external-field control over a single physical qubit to protect a quantum-gate operation. The necessary condition in this case is that the control cycle be faster than the average time between environmental interventions. Actually, in the great majority of the cases studied in the literature, methods based on DD utilize, to protect the information, pulsed fields, and it is considered that such pulses are so fast that the universe does not act over the system during their application. Nevertheless, initial attempts to use continuously-applied fields instead of pulses have appeared recently \cite{romero04, romero06, chen06}. In a previous work, we analyzed continuous decoupling during a Hadamard quantum gate, perturbed by a thermal bath of scalar bosons \cite{fffanchini07}. We showed, for that case, that the decoherence can always be efficiently reduced by applying a superposition of two external vector fields: one rotating orthogonally to the direction of the other, which remains static.

Here, we analyze the fidelity of the Hadamard quantum logic operation coupled to an environment that affects the system giving rise to three classes of errors: dephasing, bit flipping, and dissipation. We examine the feasibility of dynamical decoupling based on continuously-applied high-frequency fields, considering two types of spectral densities for the environment: ohmic and super-ohmic. Initially, we observe the fidelity of an unprotected Hadamard quantum gate coupled to a dephasing bosonic reservoir and then illustrate the ohmic and super-ohmic cases. We define an arrangement where, in the absence of protecting control fields, the quantum-gate fidelity is lower in the ohmic case than in the super-ohmic. We show that the dynamical decoupling is achieved in the super-ohmic case only when the control-field frequency is sufficiently higher than in the ohmic case. Subsequently, we consider the presence of two additional independent reservoirs, producing bit flipping and dissipation. We show, as a function of the relative coupling strengths, the relevance of bit flipping and dissipation on a phase-protected Hadamard operation. Finally, we introduce continuously-applied external fields, criteriously calculated to protect the Hadamard operation against all three independent classes of errors. 

This article is organized as follows. In Sec. II we introduce the interaction of the qubit with its environment and write down the time-local second-order master equation, describing the evolution of the reduced density matrix of the qubit in the interaction picture. In Sec. III we introduce the three independent reservoirs, describing dephasing, bit flipping, and dissipation. In Sec. IV we consider the fidelity of a Hadamard quantum gate operating under the influence of a dephasing environment and compare the ohmic and super-ohmic cases. We analyze the dynamics for all qubit initial conditions and estimate the worst fidelity with and without protection by continuous dynamical decoupling. We focus on the super-ohmic case in Sec. V and add independent bit flipping and dissipation to study the influence of these perturbations on the Hadamard evolution already protected against phase errors. In this section we also calculate the necessary control fields to eliminate all three classes of noise during the quantum-gate operation. We summarize our conclusions in Sec. VI.

\section{The Master Equation}
We start by observing that an ideal driven evolution of a qubit, free from environmental noise, is given by the action of a general time-dependent unitary transformation of the form:
\begin{eqnarray}
U(t)=I\cos\left[\alpha(t)\right]-i {\bm \sigma}\cdot{\bf \hat{u}}(t)\sin\left[\alpha(t)\right],\label{U(t)}
\end{eqnarray}
where $I$ is the $2\times2$ identity matrix, $\alpha\left(t\right)$ is an arbitrary function of time $t$, ${\bf \hat{u}}\left(t\right)$ is an arbitrary time-dependent unit vector, and ${\bm \sigma}$ is the vector matrix given by
\begin{eqnarray}
{\bm \sigma}={\bf \hat{x}}\left(
\begin{array}{cc}
0 & 1 \\
1 & 0 \\
\end{array}
\right)+{\bf \hat{y}}\left(
\begin{array}{cc}
0 & -i \\
i & 0 \\
\end{array}
\right)+{\bf \hat{z}}\left(
\begin{array}{cc}
1 & 0 \\
0 & -1 \\
\end{array}
\right).\nonumber
\end{eqnarray}
The time-dependent Hamiltonian $H_{U}(t)$, giving rise to the evolution operator in Eq. (\ref{U(t)}), is obtained by differentiation:
\begin{eqnarray}
H_{U}(t)=i\hbar\frac{dU(t)}{dt}U^{\dagger}(t)=\hbar{\bm \Omega}(t)\cdot {\bm \sigma},\label{HU}
\end{eqnarray}
with
\begin{eqnarray}
{\bm \Omega}(t)&=&\frac{d\alpha(t)}{dt}{\bf \hat{u}}(t)+\sin[\alpha (t)]\cos[\alpha (t)]\frac{d{\bf \hat{u}}(t)}{dt} \nonumber \\
&+&\sin ^{2}[\alpha (t)]{\bf \hat{u}}(t){\bf \times}\frac{d{\bf \hat{u}}(t)}{dt},\label{Omega}
\end{eqnarray}
where $\hbar$ is Planck's constant divided by $2\pi$ and $U^{\dagger}(t)$ is the Hermitian conjugate of $U(t)$.

We assume the interaction between the qubit and its environment is sufficiently weak that linear-response theory is applicable. We represent the action of the environment by
\begin{eqnarray}
H_{\rm int}= ({\bm{\mathcal L}} + {\bm{\mathcal L}}^\dagger)\cdot \bm\sigma,  \label{Hint1}
\end{eqnarray}
where ${\bm{\mathcal L}}= {\mathcal L}_1{\bf \hat{x}}+{\mathcal L}_2{\bf \hat{y}}+{\mathcal L}_3\bf \hat{z}$ is a vector operator whose components ${\mathcal L}_1$, ${\mathcal L}_2$, and ${\mathcal L}_3$ act on the environmental Hilbert space and ${\bm{\mathcal L}}^\dagger$ is the Hermitian conjugate of ${\bm{\mathcal L}}$.

The time-local, second-order master equation describing the evolution of the reduced density matrix of the qubit, in the interaction picture, is written as \cite{shibata77}:
\begin{eqnarray}
\frac{d\rho _{I}(t)}{dt}=-\int^{t}_{0}dt^{\prime} {\rm Tr}_{B}\left\{{\left[H_{I}(t),\left[H_{I}(t^{\prime}),\rho _{B}\rho _{I}(t)\right]\right]}\right\},\label{master}
\end{eqnarray}
where $H_{I}(t)$ is the interaction Hamiltonian in the interaction picture, namely,
$H_{I}(t)=U^{\dagger}(t)U^{\dagger}_{B}(t)H_{\rm int}U_{B}(t)U(t)$, with
\begin{eqnarray}
U_{B}(t)=\exp\left(-iH_{B}t\right),\label{UB}
\end{eqnarray}
where $H_{B}$ is the environmental Hamiltonian, and $U(t)$ is as in Eq. (\ref{U(t)}). Here and in the following we use units of $\hbar =1$. Equation (\ref{master}) is valid in the regime in which the strength of the coupling, expressed in frequency units, multiplied by the correlation time of the environmental operators is much lesser than unity.
Above, $\rho _{B}$ is the initial density matrix of the environment,
\begin{eqnarray}
\rho _{B}=\frac{1}{Z}\exp(-\beta H_{B}),\label{rhoB}
\end{eqnarray}
where $Z$ is the partition function given by
\begin{eqnarray}
Z={\rm Tr}_{B}\left[\exp(-\beta H_{B})\right],\label{Z}
\end{eqnarray}
$\beta =1/k_{B}T$, $k_{B}$ is the Boltzmann constant, and $T$ is the absolute temperature of the environment.

From the form of the interaction between the qubit and its environment, Eq. (\ref{Hint1}), we obtain
\begin{eqnarray}
H_{I}(t)= {\bm \Lambda }(t)\cdot [\tilde{\bm{\mathcal L }}(t) + \tilde{\bm{\mathcal L }}^\dagger(t)],\label{HI1}
\end{eqnarray}
with $\tilde{\bm{\mathcal L }}(t)=U_{B}^{\dagger}(t){\bm{\mathcal L }} U_{B}(t)$, and ${\bm \Lambda }(t)=U^\dagger(t){\bm\sigma}U(t)$. Since ${\bm \Lambda }(t)$ is just a rotation of ${\bm\sigma}$, it is convenient to write its components as
\begin{eqnarray}
{\Lambda_{\mu}(t)}= \sum_{\nu =1}^{3} R_{\mu,\nu}(t)\sigma_{\nu},\label{Lambdamu}
\end{eqnarray}
where $R_{\mu,\nu}(t)$, for $\mu,\nu=1,2,3$, are the elements of the time-dependent rotation matrix corresponding to the unitary transformation represented by $U(t)$ of Eq. (\ref{U(t)}).

By substituting Eqs. (\ref{HI1}) and (\ref{Lambdamu}) into Eq. (\ref{master}), we obtain the master equation
\begin{eqnarray}
\frac{d\rho _{I}(t)}{dt}=\sum^{3}_{\alpha,\beta=1} D_{\alpha \beta}(t)\left[\sigma _{\alpha},\rho _{I}(t)\sigma _{\beta}\right]\nonumber\\
+\sum^{3}_{\alpha,\beta=1}{D_{\alpha \beta}}^{\ast}(t)\left[\sigma _{\beta}\rho _{I}(t),\sigma _{\alpha}\right],\label{master2}
\end{eqnarray}
where we have defined
\begin{eqnarray}
D_{\alpha \beta}(t)=\sum^{3}_{\mu,\nu=1}R_{\mu,\alpha}(t) \int^{t}_{0}dt^{\prime}R_{\nu,\beta}(t^\prime )C_{\mu,\nu}(t,t^{\prime}),\label{D}
\end{eqnarray}
with
\begin{eqnarray}
\!\!\!\!\!\!\!\!\!\!\!\!\!\!\!\!C_{\mu,\nu}(t,t^{\prime})&=&\nonumber\\
& & \!\!\!\!\!\!\!\!\!\!\!\!\!\!\!\!\!\!\!\!\!\!\!\! {\rm Tr} _{B}\left[\left({\mathcal {\tilde L}}_{\mu }(t)+{\mathcal {\tilde L}}_{\mu}^\dagger(t)\right)\rho _{B}\left({\mathcal {\tilde L}}_{\nu }(t^{\prime})+{\mathcal {\tilde L}}_{\nu }^\dagger(t^{\prime})\right)\right].\label{Corr}
\end{eqnarray}
It is important to notice that Eq. (\ref{master2}) is not restricted to a specific choice of the environmental Hamiltonian. Below, we define three independent reservoirs of harmonic oscillators describing the environment, each one representing a distinct class of quantum errors.

\section{Independent Reservoirs}
We assume an environmental Hamiltonian expressed as
\begin{eqnarray}
H_{B}=\sum_{i=1}^3\sum _{k} \omega _{i,k} a^{\dagger}_{i,k}a_{i,k},\label{HB}
\end{eqnarray} 
where $\omega _{i,k}$ is the frequency of normal mode $k$ of the $i$-th independent reservoir. We take the operator ${\bm{\mathcal L}}$ as given by
\begin{eqnarray}
{\bm{\mathcal L}}=\sum_{i=1}^{3}{\bm \lambda_i}B_{i},\label{el}
\end{eqnarray}
with
\begin{eqnarray}
B_{i}=\sum_k g_{i,k}a_{i,k},\label{Bi}
\end{eqnarray}
where $g_{i,k}$ is the complex coupling constant for mode $k$ of the $i$-th reservoir with dimension of frequency,
$a_{i,k}$ is the operator that annihilates a quantum in mode $k$ of the $i$-th reservoir, and here ${\bm \lambda_i}$ is called the error vector of the $i$-th reservoir. We choose the error vectors as
\begin{eqnarray}
{\bm \lambda_1}&=&\hat{\bf x},\nonumber\\
{\bm \lambda_2}&=&\frac{\hat{\bf x}+i\hat{\bf y}}{2},\label{lambdas}\\
{\bm \lambda_3}&=&\hat{\bf z},\nonumber
\end{eqnarray}
representing bit flipping, dissipation, and phase-error, respectively. Thus, the interaction Hamiltonian, Eq. (\ref{Hint1}), is written as:
\begin{eqnarray}
H_{int}=\sum_{i=1}^3(B_i{\bm \lambda_i}+B_i^\dagger{\bm \lambda_i^{\ast}})\cdot{\bm\sigma}\label{HintB},
\end{eqnarray}
so that, by comparing Eq. (\ref{Hint1}) with Eq. (\ref{HintB}), we find that the components of ${\bm{\mathcal L}}$ in the interaction picture are given by:
\begin{eqnarray}
\mathcal {\tilde{L}}_1 (t)&=&\tilde{B}_1 (t) + \frac{1}{2} \tilde{B}_2 (t),\nonumber\\
\mathcal {\tilde{\bm L}}_2 (t) &=&i\frac{1}{2} \tilde{B}_2 (t),\label{LI}\\
\mathcal {\tilde{\bm L}}_3 (t) &=&\tilde{B}_3 (t),\nonumber
\end{eqnarray}
and, using Eqs. (\ref{UB}), (\ref{HB}), and (\ref{Bi}),
\begin{eqnarray}
\tilde{B}_i(t)=U_B^\dagger(t) {B}_i U_B(t)=\sum_k g_{i,k}a_{i,k}\exp(-i\omega_{i,k}t).\label{Bitil}
\end{eqnarray}

With these considerations, we calculate the reservoir correlation function of Eq. (\ref{Corr}), $C_{\mu,\nu}(t,t^{\prime})$, using Eqs. (\ref{rhoB}), (\ref{Z}), (\ref{LI}), and (\ref{Bitil}):
\begin{eqnarray}
\!\!\!\!\!\!\!\! \mathcal{I}_{i,j}^{(1)}(t-t^\prime)&\equiv &{\rm Tr} _{B}\left[ \tilde{B}_{i}(t)\rho _{B} {\tilde B}_{j}^\dagger(t^{\prime})\right]\nonumber\\
& &\!\!\!\!\!\!\!\!\!\!\!\!\!\!\!\! =\delta_{ij}\sum_k\left|g_{i,k}\right|^2 n_{i,k}\exp[-i\omega_{i,k}(t-t^\prime)]\label{I1}
\end{eqnarray}
and
\begin{eqnarray}
\!\!\!\!\!\!\!\! \mathcal{I}_{i,j}^{(2)}(t-t^\prime)&\equiv &{\rm Tr} _{B}\left[ \tilde{B}_{i}^\dagger(t)\rho _{B} {\tilde B}_{j}(t^{\prime})\right]\nonumber\\
& &\!\!\!\!\!\!\!\!\!\!\!\!\!\!\!\!\!\!\!\! =\delta_{ij}\sum_k\left|g_{i,k}\right|^2 (n_{i,k}+1) \exp[i\omega_{i,k}(t-t^\prime)],\label{I2}
\end{eqnarray}
where $n_{i,k}$ is the average occupation number of mode $k$ of the $i$-th reservoir:
\begin{eqnarray}
n_{i,k}=\frac{1}{\exp(\beta \omega_{i,k})-1}.\label{nik}
\end{eqnarray}
Here $\mathcal{I}_{i,j}^{(1)}(t-t^\prime)$ and $\mathcal{I}_{i,j}^{(2)}(t-t^\prime)$ determine how much of the recent past history of the components of each coupling, effectively, contributes to the time average in Eq. ({\ref{D}}). Hence,
Eq. (\ref {master2}) is not restricted by a Markovian approximation.
In the limit in which the number of modes per unit frequency becomes infinite, we interpret the summations in Eqs. (\ref{I1}) and (\ref{I2}) as integrals:
\begin{eqnarray}
\mathcal I_{i,j}^{(1)}(t)=\int_0^\infty d\omega J_i(\omega)n_i(\omega)\exp(-i\omega t)\label{I12}
\end{eqnarray}
and
\begin{eqnarray}
\mathcal I_{i,j}^{(2)}(t)=\int_0^\infty d\omega J_i(\omega)\exp(i\omega t)[n_i(\omega)+1],\label{I22}
\end{eqnarray}
with $n_{i}(\omega)$ being the continuous-frequency version of Eq. (\ref{nik}):
\begin{eqnarray}
n_{i}(\omega)=\frac{1}{\exp(\beta \omega)-1},\label{niomega}
\end{eqnarray}
and we define the spectral density as 
\begin{eqnarray}
J_i(\omega)=\eta_i\frac{\omega^{s_i}}{{\omega_c}^{s_i-1}}\exp(-\omega/{\omega_c}),\label{Ji}
\end{eqnarray}
where $\eta_i$ is a dimensionless constant, proportional to the coupling strength of the system to the $i$-th reservoir, $s_i$ defines the environment spectral density of the $i$-th reservoir as ohmic $(s_i=1)$ or super-ohmic $(s_i>1)$, and $\omega_c$ is a cut-off frequency. Hence, using Eqs. (\ref{niomega}) and (\ref{Ji}), the integrals of Eqs. (\ref{I12}) and (\ref{I22}) are calculated, giving:
\begin{eqnarray}
\int_0^\infty d\omega J_i(\omega)\exp(-i\omega t)=\eta_i {\omega_c}^2 \frac{s_i!}{\left[1+i{\omega_c} t\right]^{s_i+1}},\label{int1}
\end{eqnarray}
and
\begin{eqnarray}
 \lefteqn{ \int_0^\infty d\omega J_i(\omega)n_i(\omega)\exp(-i\omega t)=}\nonumber\\
 & &=\eta_i {\omega_c}^2\sum_{n=0}^\infty\	\frac{s_i!}{\left[1+i{\omega_c} t+\beta{\omega_c}(n+1) \right]^{s_i+1}}.\label{int2}
\end{eqnarray}
With these results, thus, we calculate $\mathcal{I}_{i,j}^{(1)}(t)$, that is given by Eq. (\ref{int2}), and $\mathcal{I}_{i,j}^{(2)}(t)$, that is given by $\mathcal{I}_{i,j}^{(1)}(t)^\ast$ plus Eq. (\ref{int1}).

\section{Protecting a Hadamard Quantum Gate from Dephasing}
In this section we analyze the fidelity of a Hadamard quantum gate protected against dephasing by a suitable continuously-applied control field. We compare the cases of ohmic and super-ohmic spectral densities. The total Hamiltonian, in the Schr\"{o}dinger picture, is given by
\begin{eqnarray}
H(t)=H_{U}(t)+H_{int}+H_{B},\nonumber
\end{eqnarray}
where $H_{U}(t)$ is to be calculated according to Eq. (\ref{HU}) once we determine $U(t)$, $H_{B}$ and $H_{int}$ are given by Eqs. (\ref{HB}) and (\ref{HintB}), respectively. For dephasing errors only, $B_{1}=0$ and $B_{2}=0$ and, in view of Eqs. (\ref{el}), (\ref{lambdas}), and (\ref{HintB}), in this section we use
\begin{eqnarray}
H_{int}=(B_3+B_3^\dagger)\sigma_{z}.\label{dephasing}
\end{eqnarray} 

Because we intend to realize a quantum logic operation simultaneously with the protection from errors, we proceed as in Ref. \cite{fffanchini07} and split $H_{U}$ into two terms:
\begin{eqnarray}
H_{U}(t) = H_0(t)+H_c(t),\label{HU2}
\end{eqnarray}
where $H_0(t)$ is to produce the quantum-gate result after a certain time interval, while $H_c(t)$ counteracts the perturbing action of the environment.

According to the general prescription for dynamical decoupling \cite{facchi05}, the unitary operator $U_c(t)$, corresponding to the control Hamiltonian $H_c(t)$, is to be periodic and our choice is:
\begin{eqnarray}
U_c(t)=I\cos(2n\pi t/\tau)-i\sigma_{x}\sin(2n\pi t/\tau),\label{uc}
\end{eqnarray}
so that it satisfies, using Eq. (\ref{dephasing}),
\begin{eqnarray}
\int_0^{t_c}dtU_c^\dagger(t) H_{int} U_c(t)=0,\nonumber
\end{eqnarray} 
where $t_c < \tau$ is the period of $U_c(t)$, Eq. (\ref{uc}), with $\tau$ being the time for completing the gate operation. Here, for convenience, we choose $\tau$ as an integer multiple of $t_c$, that is, $\tau=nt_c$, such that $U_c(\tau) = I$.

In the picture obtained using the unitary transformation $U_c(t)$ as given by Eq. (\ref{uc}), we choose $H_0(t)$ such that it produces the intended Hadamard gate at instant $\tau$:
\begin{eqnarray}
U_c^\dagger(t)H_0(t)U_c(t)=\frac{\pi}{2\tau}\frac{\sigma_{x}+\sigma_{z}}{\sqrt{2}}.\label{H0}
\end{eqnarray}
Since $U_c^\dagger(t)H_0(t)U_c(t)$ is taken, in Eq. (\ref{H0}), as time independent, the unitary evolution operator associated with this quantity is given by
\begin{eqnarray}
U_0(t)=I\cos(\pi t/2\tau) - i\frac{\sigma_{x}+\sigma_{z}}{\sqrt{2}}\sin(\pi t/2\tau).\label{U0}
\end{eqnarray}
We define $U(t)$ as the composed unitary operator
\begin{eqnarray}
U(t) = U_c(t)U_0(t),\label{U}
\end{eqnarray}
noticing that, at $t = \tau$, $U_c(\tau)=I$. Thus, at time $\tau$, we obtain, up to an irrelevant global phase factor, the desired Hadamard quantum gate:
\begin{eqnarray}
U(\tau)=- i\frac{\sigma_{x}+\sigma_{z}}{\sqrt{2}}.\nonumber
\end{eqnarray}

Thus, multiplying Eqs. (\ref{uc}) and (\ref{U0}) gives $U(t)$ for the particular case of a Hadamard gate protected against dephasing and, comparing with Eq. (\ref{U(t)}), we obtain:
\begin{eqnarray}
\cos[\alpha(t)]&=&-\sin(2n\pi t/\tau)\sin(\pi t/2\tau)/\sqrt{2}\nonumber\\
& &+\cos(2n\pi t/\tau)\cos(\pi t/2\tau),\label{cos}\\
{\bf \hat{u}}(t)\sin[\alpha(t)]&=&-{\bf \hat{y}}\sin(2n\pi t/\tau)\sin(\pi t/2\tau)/\sqrt{2}\nonumber\\
& &+({\bf \hat{x}}+{\bf \hat{z}})\cos(2n\pi t/\tau)\sin(\pi t/2\tau)/\sqrt{2}\nonumber\\
& &+{\bf \hat{x}}\sin(2n\pi t/\tau)\cos(\pi t/2\tau)\label{usin}.
\end{eqnarray}
The explicit form of the Hamiltonian $H_U$ of Eq. (\ref{HU2}) is the one already given by Eq. (\ref{HU}), $H_U ={\bm \Omega}(t)\cdot{\bm\sigma}$, where the applied external field is calculated from Eqs. (\ref{cos}) and (\ref{usin}) according to the prescription of Eq. (\ref{Omega}):
\begin{eqnarray}
&&{\bm \Omega}(t)=\frac{\pi}{\tau}\left(2n+\frac{1}{2\sqrt{2}}\right){\bf \hat{x}}\nonumber \\
&&-\frac{\pi}{2\sqrt{2}\tau}\left[{\bf\hat{y}}\sin\left(\frac{4n\pi t}{\tau}\right) - {\bf \hat{z}}\cos\left(\frac{4n\pi t}{\tau}\right) \right].\label{field1}
\end{eqnarray}
The first term of Eq. (\ref{field1}) is a static field along the direction that is perpendicular to the error vector, and the other two terms give a rotating field perpendicular to the direction of the static field, as already shown in Ref. \cite{fffanchini07} for an arbitrary error vector in the case of a single boson field.

\begin{figure}
\includegraphics[width=8.0cm]{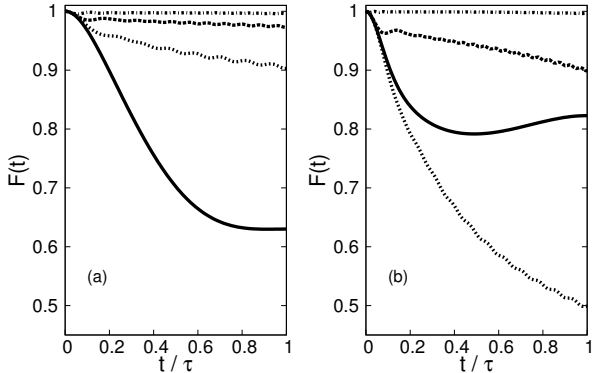}
\caption{\label{figure1} Fidelities as functions of time, obtained by solving Eq. (\ref{master2}) numerically, as described in the text, exemplifying protection against phase-errors during a Hadamard operation, in the cases of ohmic (a) and super-ohmic (b) spectral densities. The solid curve represents the fidelity when $H_{c}=0$, while the dotted, the dashed, and the dot-dashed curves represent, respectively, the cases of $n=2$, $n=3$, and $n=5$ for (a), and $n=3$, $n=5$, and $n=15$ for (b).}
\end{figure}
\begin{figure}
\includegraphics[width=8cm]{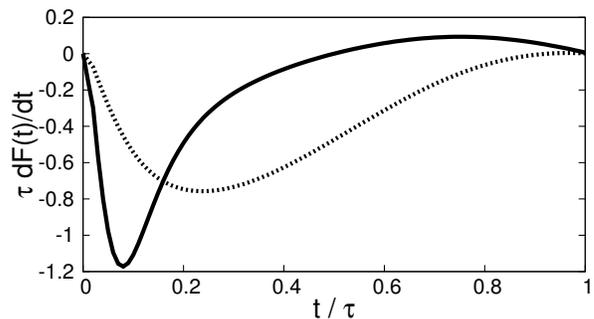}
\caption{\label{figure2} Derivatives of the fidelities of an unprotected Hadamard quantum gate in the ohmic (dashed curve) and super-ohmic (solid curve) cases of environmental spectral densities.}
\end{figure}

For the particular case of this section, that is, a Hadamard quantum gate protected against dephasing and not subjected to any other source of noise, we solve Eq. (\ref{master2}) numerically. In Eq. (\ref{Ji}), for $i=3$, we take $\omega_{c}\tau=2\pi$ and $\eta_{3}=1/16$, with $\tau=10^{-10}$s. Assuming $T=0.25$K and the initial condition $\rho_{I}(0)=I/2+\sigma_{x}/2$, we calculate, as shown in Fig. \ref{figure1}, the fidelity as a function of time, ${F}(t)={\rm Tr}[\rho_{I}(t)\rho_{I}(0)]$,  for two spectral-density cases: ohmic, with $s_{3}=1$, and super-ohmic, with $s_{3}=3$. When $H_{c}=0$ the calculation gives ${F}(\tau)={\rm Tr}[\rho_{I}(\tau)\rho_{I}(0)]\approx0.6299$ for the ohmic case and ${F}(\tau)\approx0.8227$ for the super-ohmic case. Now, if we turn on the control field of Eq. (\ref{field1}), the fidelity becomes, for an ohmic spectral density, ${F}(\tau)\approx0.9956$ for $n$ as low as $5$. However, in the super-ohmic case, to obtain a fidelity of the same order, $n$ must be higher than $14$, at least, for $n=15$ gives ${F}(\tau)\approx0.9960$. Hence, even though a super-ohmic reservoir causes less damage to the processing of quantum information than an ohmic reservoir, it requires a higher-frequency field to protect the intended quantum evolution from the environmental noise. To understand this result, Fig. \ref{figure2} shows the derivatives of the fidelities, without the control-field protection, for the ohmic ($s_{3}=1$) and super-ohmic ($s_{3}=3$) cases. We observe, as expected, a faster decoherence, up to $t/\tau\approx0.1$, in the super-ohmic case than in the ohmic.

Now we investigate the fidelity of the Hadamard operation for all initial conditions corresponding to a pure state of the qubit. Therefore, if $\rho(0)$ is pure, then
\begin{eqnarray}
\rho(0)=\frac{1}{2}{\bf I}+\frac{1}{2}{\bf \hat{r}}\cdot\bm\sigma,\label{rho0}
\end{eqnarray}
where ${\bf \hat{r}}$ is a real unit vector written in terms of the two spherical polar angles:
\begin{equation}
{\bf \hat{r}}={\bf\hat{x}}\sin\theta\cos\varphi+{\bf\hat{y}}\sin\theta\sin\varphi+{\bf\hat{z}}\cos\theta.\nonumber
\end{equation}
\begin{figure}
\includegraphics[width=8.5cm]{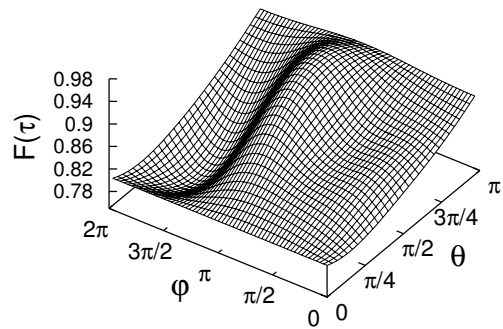}
\caption{\label{figure3} Fidelity (at $t=\tau$) of a Hadamard quantum gate as a function of the initial conditions. The initial density operator, $\rho(0)$, is given by Eq. (\ref{rho0}). The qubit is coupled to a super-ohmic reservoir ($s_{3}=3$) that causes phase-errors and no protecting external fields are present.}
\end{figure}
Thus, using the numbers as before, in the super-ohmic case ($s_{3}=3$), when there is only dephasing due to the environment and the control fields are off, Fig. \ref{figure3} shows the fidelity for all possible pure-state initial conditions given by Eq. (\ref{rho0}). If the protecting fields of Eq. (\ref{field1}) are present, at gate completion ($t=\tau$) we obtain the results of Fig. \ref{figure4} with $n=25$.
\begin{figure}
\includegraphics[width=8.5cm]{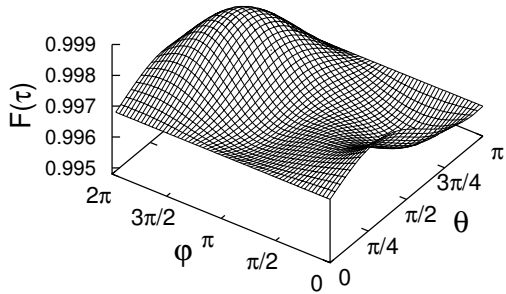}
\caption{\label{figure4} Fidelity (at $t=\tau$) of a Hadamard quantum gate as a function of the initial conditions. The initial density operator, $\rho(0)$, is given by Eq. (\ref{rho0}). The qubit is coupled to a super-ohmic reservoir ($s_{3}=3$) that causes phase-errors and, in this case, the control fields of Eq. (\ref{field1}) with $n=25$ are continuously applied to protect the gate operation.}
\end{figure}
We notice that, comparing Figs. \ref{figure3} and \ref{figure4}, with $n$ as low as $25$ in Eq. (\ref{field1}) we already can increase the quantum-gate fidelity significantly even in the situation of a super-ohmic environment. For the lowest fidelities of both figures we find ${F_{\rm min}}(\tau)\approx 0.7525$ for the unprotected gate of Fig. \ref{figure3} and ${F_{\rm min}}(\tau)\approx 0.9951$ for the protected dynamics of Fig. \ref{figure4}.

\section{Protecting a Hadamard quantum gate from independent dephasing, bit flipping, and dissipation}
In the previous section, we have shown how decoherence due to dephasing can be efficiently reduced during logical operations by applying a superposition of two external vector fields: one rotating orthogonally to the direction of the other, which remains static \cite{fffanchini07}. In this section we maintain the dephasing reservoir with a super-ohmic spectral density, but include other two independent reservoirs, corresponding to bit flipping and dissipation. The corresponding error vectors are given by Eq. (\ref{lambdas}).

\begin{figure}
\includegraphics[width=8cm]{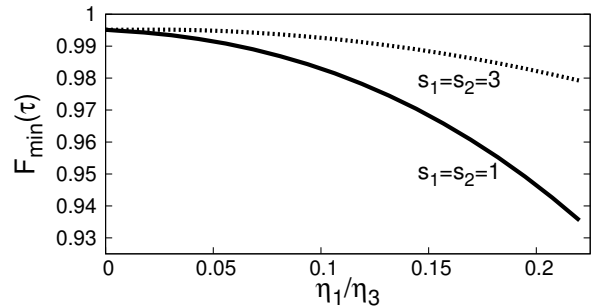}
\caption{\label{figure5} Lowest fidelity of all those corresponding to the initial conditions of Eq. (\ref{rho0}), as a function of $\eta_{1}/\eta_{3}$ with $\eta_{2}=\eta_{1}$. Here, the gate is protected from dephasing only, but is subjected to independent bit flipping and dissipation.}
\end{figure}

We start this section by assuming that the applied external fields are those of Eq. (\ref{field1}), designed to protect the Hadamard gate from phase noise only. For simplicity, we take the spectral densities of the additional reservoirs to be either both ohmic or super-ohmic, with $\eta_{1}=\eta_{2}$ in Eq. (\ref{Ji}). In Fig. \ref{figure5} we show the results of selecting the lowest fidelity of all those corresponding to the initial conditions of Eq. (\ref{rho0}), as a function of $\eta_{1}/\eta_{3}$. We notice that for $\eta_{1}/\eta_{3}<0.01$ the final result changes insignificantly and protecting from phase errors only is enough to achieve high fidelity during the time interval $\tau$. This figure also points out that bit flipping and dissipation are more relevant in the ohmic case than in the super-ohmic.

To achieve efficient simultaneous reduction of dephasing, bit flipping, and dissipation, we modify Eq. (\ref{uc}) by redefining $U_{c}(t)$:
\begin{eqnarray} 
U_c(t)=U_{cx}(t)U_{cz}(t),\label{u2}
\end{eqnarray}
with
\begin{eqnarray} 
U_{cx}(t)&=&I\cos(2n\pi t/\tau)-i{\sigma_x}{\sin(2n\pi t/\tau)},\label{ucx}\\
U_{cz}(t)&=&I\cos(2m\pi t/\tau)-i{\sigma_z}{\sin(2m\pi t/\tau)},\label{ucz}
\end{eqnarray}
where $m$ is an integer different from $n$. The operator $U_{cx}(t)$ corresponds to protection against dephasing, as in Sec. IV, and the operator $U_{cz}(t)$ is associated with the protection from bit flipping and dissipation. The unitary evolution $U(t)$ is again defined as in Eq. (\ref{U}), but now with $U_c(t)$ given by Eqs. (\ref{u2}), (\ref{ucx}), and (\ref{ucz}).

The control Hamiltonian $H_{U}(t)$ is obtained from Eqs. (\ref{HU}), (\ref{U}), (\ref{u2}), (\ref{ucx}), and (\ref{ucz}), with the components of the control field ${\bm \Omega}(t)$ given by
\begin{widetext}
\begin{eqnarray}
{\Omega}_x(t)&=&\frac{\pi}{\tau}\left[2n+\frac{1}{2\sqrt{2}}\cos\left(\frac{4m\pi t}{\tau}\right)\right],\label{field3x}\\
{\Omega}_y(t)&=&-\frac{\pi}{\tau}\left[2m+\frac{1}{2\sqrt{2}}\right]\sin\left(\frac{4n\pi t}{\tau}\right) + \frac{\pi}{2\sqrt{2}\tau}\cos\left(\frac{4n\pi t}{\tau}\right)\sin\left(\frac{4m\pi t}{\tau}\right),\label{field3y}\\
{\Omega}_z(t)&=&\frac{\pi}{\tau}\left[2m+\frac{1}{2\sqrt{2}}\right]\cos\left(\frac{4n\pi t}{\tau}\right) + \frac{\pi}{2\sqrt{2}\tau}\sin\left(\frac{4n\pi t}{\tau}\right)\sin\left(\frac{4m\pi t}{\tau}\right). \label{field3z}
\end{eqnarray}
\end{widetext}
Using this new field we can protect the Hadamard gate from simultaneous dephasing, bit flipping and dissipation. As an illustration of this protection, we calculate the lowest fidelity of all those corresponding to the initial conditions of Eq. (\ref{rho0}) in the case of super-ohmic dephasing as specified in Sec. IV, but when the bit flipping and dissipation both have either ohmic or super-ohmic spectral densities. Using $\eta_{1}=\eta_{2}=0.2$, $n=25$, $m=10$, and the remaining numbers as in Sec. IV, we obtain ${F_{\rm min}}(\tau)\approx0.9938$ in the ohmic case and 
${F_{\rm min}}(\tau)\approx0.9962$ in the super-ohmic case. These fidelities are to be compared with their respective results, $\sim \!\!\!\; 0.9466$ and $\sim \!\!\!\; 0.9822$, when all three reservoirs are present, but only dephasing is protected with the field of Eq. (\ref{field1}). These increases are, respectively, $5\%$ and $1.5\%$, reaffirming the results of Sec. IV that protection in a super-ohmic environment requires higher frequencies than in an ohmic.

\section{CONCLUSION}

We analyze a dynamical-decoupling method based on continuously-applied fields to protect a Hadamard quantum gate from a perturbing environment. We find that, for a reservoir with an ohmic spectral density, the gate fidelity can be lower than in the super-ohmic case, but it requires lower field frequencies to decouple the qubit dynamics from the environment. We show that this characteristic situation occurs because, when the system interacts with a super-ohmic reservoir, the fidelity decreases drastically already at the beginning of the quantum logic operation, but, soon, the decrease saturates as the evolution progresses. For an ohmic reservoir, however, the fidelity decreases during all the gate operation, but smoothly. This peculiarity makes the ohmic reservoir more harmful to the quantum gate operation than the super-ohmic, but easier to be protected against.

We also study the relevance of bit flipping and dissipation to the gate fidelity when it is protected from dephasing only. We show how robust this partial protection is against these other perturbations. Finally, we calculate an efficient field arrangement, given by Eqs. (\ref{field3x}), (\ref{field3y}), and (\ref{field3z}), capable of protecting simultaneously the gate operation from dephasing, bit flipping and dissipation.

\begin{acknowledgments}
This work has been supported by Funda\c{c}\~{a}o de Amparo \`{a} Pesquisa do Estado de S\~{a}o
Paulo, Brazil, project number 05/04105-5 and the Millennium Institute for Quantum Information -- Conselho Nacional de Desenvolvimento Cient\'{\i}fico e Tecnol\'{o}gico, Brazil.
\end{acknowledgments}

\end{document}